\documentclass[showpacs,preprintnumbers,amsmath,amssymb,pra,twocolumn]{revtex4}
\usepackage{stmaryrd}
\usepackage{amssymb}
\usepackage{amsfonts}
\usepackage{mathrsfs}
\usepackage{graphicx}
\usepackage{dcolumn}
\usepackage{bm}
\usepackage{verbatim}
\begin{document}
\title{ Entanglement-assisted local operations and classical
communications conversion in the quantum critical systems}

\author{
Jian Cui, Jun-Peng Cao, Heng Fan } \affiliation{ Institute of
Physics, Chinese Academy of Sciences, Beijing National Laboratory
for Condensed Matter Physics, Beijing 100190, China }
\date{\today}

\begin{abstract}
Conversions between the ground states in quantum critical systems
via entanglement-assisted local operations and classical
communications (eLOCC) are studied. We propose a new method to
reveal the different convertibility by local operations when a
quantum phase transition occurs. We have studied the ground state
local convertibility in the one dimensional transverse field Ising
model, XY model and XXZ model. It is found that the eLOCC
convertibility sudden changes at the phase transition points. In
transverse field Ising model the eLOCC convertibility between the
first excited state and the ground state are also distinct for
different phases. The relation between the order of quantum phase
transitions and the local convertibility is discussed.

\end{abstract}

 \maketitle

\section{Introduction}
The recent development in quantum information processing (QIP)
\cite{nature review} has provided much insight into the quantum
many-body physics \cite{REVIEWS OF MODERN PHYSICS}. In particular,
using the ideas from QIP to investigate quantum phase transitions
has drawn vast attentions and has been successful in detecting a
number of critical points of great interest. For instance,
entanglements measured by concurrence, negativity, geometric
entanglement and von Neumann entropy are studied in several critical
systems \cite{von Neumann entropy,Nielsen ising,Osterloh
nature,LAW,TCW}. It was found that the von Neumann entropy diverges
logarithmically at the critical point \cite{von Neumann entropy},
and the concurrence shows a maximum at the quantum critical points
of transverse field Ising model and XY model \cite{Nielsen ising}.
The second order phase transition in the XY model can also be
determined by the divergence of the concurrence derivative or the
negativity derivative with respect to the external field parameter
\cite{Osterloh nature,LAW}. Apart from entanglement, other concepts
in quantum information such as mutual information and quantum
discord which feature some singularities at critical points were
also found to be useful in detecting quantum phase transitions
\cite{QD,jiancui}. Recent studies in entanglement spectrum can be
applied to the detection of quantum phase transitions \cite{lihui
Dillenschneider hongyao,xiaogangwen}, too. At the same time,
fidelity as well as the fidelity susceptibility of the ground state
also works well in exploring numerous phase transition points in
various critical systems \cite{Quan polozanardi gushijianreview
Buonsnate zanadiPRL zhou,gushijian,quanPRE}. These methods have
achieved great success in understanding the deep nature of the
different phase transitions, especially the structure of the
correlations revolved \cite{AA,LCV}.

However, in the previous studies although the concepts from quantum
information were investigated, they were not fully explored, because
they were mealy used as some common order parameters, and some
important operational properties were missing.
 In this paper, from a new point of view, we reveal the operational properties of the critical
 systems by fully studying the ground state R\'{e}nyi entropy and show that
the operational property sudden changes at the quantum phase
transition point, so that we could put forward a new proposal to
investigate the quantum phases and quantum phase transitions. To be
specific, we investigate a many-body system whose Hamiltonian is
$H(\lambda)$ with a variable parameter $\lambda$. The system has a
critical point at $\lambda=\lambda_c$, which separates two quantum
phases. We examine the possibility for the ground state of
$H(\lambda)$ to convert into the ground state of $H(\lambda')$ by
entanglement-assisted local operations and classical communications
(eLOCC). If we can find a method to convert them, we say there is
local convertibility between them, otherwise there is no local
convertibility. Our motivation is that from the R\'{e}nyi entropy
interceptions of different states we can get the knowledge of their
local convertibility, and this local convertibility is different for
ground states from different phases. Thus, phase transitions can be
distinct observed. Besides the operational meaning, our proposal has
other advantages in that it is easy to moderate the accuracy, and
the quantum phase transition points can be detected for
 small-sized system. In addition, we do not need a priori
knowledge of the order parameters nor the symmetry.
 As we are revealing the local operation properties of different quantum phases,
  this paper is essentially concerned with deterministic conversion
of a single copy of state, which would be different from the
stochastic results \cite{SLOCC} and asymptotic results
\cite{asymptotic}.

The paper is organized as follows. In section II, we introduce some
local conversions as well as their necessary and sufficient
conditions, especially the eLOCC conversion, which is mainly focused
on in this paper. In section III, we study the ground state local
convertibility in the one dimensional spin half transverse field
Ising model, XY model and XXZ model. In particular, for the
transverse field Ising model we show how to determine the critical
point numerically with small-sized systems by the finite size
scaling analysis, and we also investigate the local convertibility
between the ground state and the corresponding first excited state
for the Ising model. In section IV we give some conclusions and
discussions.

\section{Local convertibility of two pure states}
In this section, we introduce the notion of local convertibility and
give the necessary and sufficient conditions for the local
convertibility used in this paper. Generally, local convertibility
concerns the following question: given two quantum states, is that
possible to convert one to the other merely using local operations
(without global operations)? If it is possible, we say there is
local convertibility between the states. Otherwise, there is no
local convertibility. The answer to this question is related to the
comparisons between the entanglements of the two states. A measure
of entanglement which does not increase in the process of LOCC
besides other basic conditions is defined as entanglement monotone
\cite{monotone}. Entanglement monotone is not unique. Different
entanglement monotones reflect certain aspects of the entanglement
and could have inequable usages in QIP. In particular, local
convertibilities within different kinds of allowed local operations
have different entanglement monotones to compose the necessary and
sufficient conditions.

The most basic local conversion is LOCC, which is also the most
natural restriction in quantum information processing. It has
fabulous applications in several fundamental problems in QIP, such
as teleportation \cite{teleportation}, quantum states discrimination
\cite{QSD}, testing the security of hiding bit \cite{hiding bit} and
quantum channel corrections \cite{channel correction}. By Schmidt
decomposition, a bipartite pure quantum state can be written as $
|\Psi \rangle _{AB}=\sum _{k=1}^d\sqrt {\lambda _k}|kk\rangle
_{AB}$, where $d$ is the rank of the reduced density operator $\rho
_{A(B)}=Tr_{B(A)}\left( |\Psi \rangle \langle \Psi |\right) $, and
$\{ \lambda _k\} _{k=1}^d$ are the Schmidt coefficients in
descending order. They satisfy the positive and normalization
conditions, i.e., $\lambda_k>0$ for all $1\leq k\leq d$, and $\sum
_k\lambda _k=1$. For a given bipartition of a real physical system
all the knowledge of the ground state is contained in the Schmidt
coefficients \cite{Poilblanc}. It is pointed out that quantities $\{
\sum _{k=l}^{d}\lambda _k\} _{l=1}^{d}$ constitute a set of
entanglement monotones \cite {vidal}. For two bipartite pure states
$|\Psi \rangle $ and $|\Psi '\rangle =\sum _{k=1}^d\sqrt {\lambda
'_k}|kk\rangle _{AB}$, if the majorization relations
\begin{eqnarray}
\sum _{k=l}^d\lambda _k\ge \sum _{k=l}^d\lambda '_k
\end{eqnarray}
are satisfied for all $1\leq l\leq d$, state $|\Psi \rangle $ can be
converted with $100\% $ probability of success to $|\Psi '\rangle $
by LOCC \cite{Nielsen}. Otherwise these two states are incomparable,
i.e., neither can state $|\Psi \rangle $ be converted to $|\Psi
'\rangle $ nor can state $|\Psi '\rangle $ be converted to $|\Psi
\rangle $ by LOCC. Thus, the majorization relations provide a
necessary and sufficient condition for the local convertibility with
LOCC. In this sense, $\{\sum _{k=l}^d\lambda _k\} _{l=1}^{d}$ is a
minimal and complete set of entanglement monotones for LOCC.

 Another useful local conversion
which is also the most powerful one is eLOCC, which is also called
entanglement catalyst. Entanglement catalyst in LOCC is such a
phenomenon that there are some pure states $|\Psi\rangle_{AB}$ and
$|\Psi\rangle_{AB}^{\prime}$ that they cannot be converted to each
other by LOCC alone, because they violate the majorization theorem.
However, when the two local parties share certain kind of ancillary
entanglement, which is labeled as $|\phi\rangle$, the state with
larger entanglement can be converted to the other state by LOCC and
the ancillary state does not change after the conversion just like
the catalyst in chemistry reactions \cite{jonathanplenio}, i.e.,
$|\Psi\otimes\phi\rangle\rightarrow|\Psi^{\prime}\otimes\phi\rangle$.
For example, $|\Psi\rangle=\sqrt{0.4} |00\rangle+\sqrt{0.4}
|11\rangle+\sqrt{0.1} |22\rangle+\sqrt{0.1} |33\rangle$, and
$|\Psi^{\prime}\rangle=\sqrt{0.5} |00\rangle+\sqrt{0.25}
|11\rangle+\sqrt{0.25} |22\rangle+\sqrt{0} |33\rangle$. It can be
 easily checked that
$\lambda_2+\lambda_3+\lambda_4>\lambda_2^{\prime}+\lambda_3^{\prime}+\lambda_4^{\prime}$,
but $\lambda_3+\lambda_4<\lambda_3^{\prime}+\lambda_4^{\prime}$,
therefore, neither state can be converted to the other with
certainty, i.e., $|\Psi\rangle\nrightarrow|\Psi'\rangle$ and
$|\Psi'\rangle\nrightarrow|\Psi\rangle$. Whereas if
$|\phi\rangle=\sqrt{0.6}|44\rangle+\sqrt{0.4}|55\rangle$ is applied,
the Schmidt coefficients for the product state
$|\Psi\rangle|\phi\rangle$ and $|\Psi^{\prime}\rangle|\phi\rangle$
in decreasing order are
$\Lambda=\{0.24,0.24,0.16,0.16,0.06,0.06,0.004,0.04\}$ and
 $\Lambda'=\{0.30,0.20,0.15,0.15,0.10,0.10,0.00,0.00\}$, such that
$\sum_{k=l}^8\Lambda_k\geq\sum_{k=l}^8\Lambda_k^{\prime}, \forall
1\le l \le 8$. As a result, $|\Psi\rangle|\phi\rangle$ can be
converted to $|\Psi^{\prime}\rangle|\phi\rangle$ with certainty by
LOCC.  We can see that the LOCC with ancillary assisted-entanglement
(eLOCC) actually enlarges the previous Hilbert space and is a
generalized version of LOCC.

The eLOCC conversion can be determined by the R\'{e}nyi entropy
\cite{Renyi}, which is defined as
\begin{eqnarray}
S_{\alpha }(\rho _A)=\frac {1}{1-\alpha }\log Tr\rho _A^{\alpha }.
\end{eqnarray}
In particular, when $\alpha\rightarrow 1$, it tends to the von
Neumann entropy. R\'{e}nyi entropy was extensively studied in spin
chain systems \cite{spinchainrenyi}. States $|\Psi \rangle _{AB}$
can be converted to $|\Psi ' \rangle _{AB}$ by eLOCC iff their
R\'{e}nyi entropies satisfy $S_{\alpha }(\rho _A)\ge S_{\alpha
}(\rho '_A)$ for all $\alpha$ \cite{necessaryandsufficient}.
 So in the graph of R\'{e}nyi entropy versus $\alpha$, if the R\'{e}nyi
 entropies of states $|\Psi \rangle _{AB}$ and $|\Psi '\rangle _{AB}$ intercept, these two states are incomparable, see Fig.1 (right). On the
other hand, if there is no interception, the state with larger
entanglement can be locally converted to the other state, see Fig.1
(left). Therefore, the R\'{e}nyi entropies are a minimal and
complete set of entanglement monotones for eLOCC. In the following,
we focus on the local convertibility within eLOCC.

\begin{figure}
\includegraphics[height=4cm,width=\linewidth]{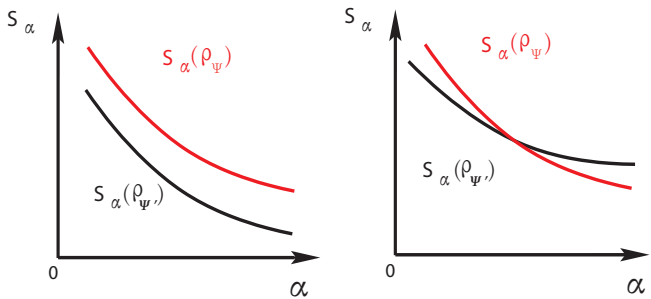}
\caption{\label{fig1} (Color online).R\'{e}nyi entropies of two
bipartite states $|\Psi \rangle $ and  $|\Psi^{\prime} \rangle $ may
have two types of behavior: they intercept or not. For interception
case (right), the two states cannot be locally converted to each
other.
 For non-interception
case (left), $|\Psi \rangle $ can be locally converted to
$|\Psi^{\prime} \rangle $. }
\end{figure}

Now we consider the R\'{e}nyi entropy in quantum critical systems.
The wave functions of ground states from different quantum phases
are drastically distinct. When quantum phase transition occurs, the
properties of the ground state must change abruptly \cite{quantum
phase transition}. We argue that as part of this general feature of
quantum phase transition, the interception of ground state R\'{e}nyi
entropies as well as the local conversion property of the ground
state will change at the critical point, and the different quantum
phases boundaries can be determined by the R\'{e}nyi entropy. By
carefully analyzing the behavior of R\'{e}nyi entropy, we find two
possible
 cases: (i) Please see Table I (left). In phases I, the R\'{e}nyi
entropies of any two different ground states intercept, while in
phases II, any two different ground states do not intercept. And the
R\'{e}nyi entropies of two states from different phases will
intercept. That means for any two states in phase II the one with
larger entanglement can be converted to the other one via eLOCC,
while in phase I any ground state cannot convert to other states via
eLOCC, as a result global operation is necessary in phase I. The
corresponding examples for this type are the transverse field Ising
model and XY model. (ii) Please see Table I (right). The ground
states R\'{e}nyi entropies do not intercept with others in the same
phase, but they intercept in the different phases. That means the
ground state can be converted through eLOCC within the same phase.
However, the ground state cannot be converted locally into the
different phases. The corresponding example is XXZ model. These two
scenarios can be used to detect quantum phase transitions.

\begin{table}
\caption{\label{tab0} Interceptions of the ground states R\'{e}nyi
entropies, where $\times$ means R\'{e}nyi entropies are intercepted
and N means the non-interception. The left table is for case (i)
where the phase boundary can be obtained along the diagonal
elements. The right table corresponds to the case (ii) where the
phase boundary can be obtained with the help of the anti-diagonal
elements.}
\begin{normalsize}  
\begin{tabular}{|c|c|c|}
\hline   &    phase I &    phase II \\
\hline phase I&   $\times$ & $\times$\\
\hline phase II&  $\times$ &    N\\
\hline
\end{tabular}
\begin{tabular}{|c|c|c|}
\hline   &    phase I &    phase II \\
\hline phase I&   N & $\times$  \\
\hline phase II&  $\times$ & N\\
\hline
\end{tabular}
\end{normalsize}
\end{table}

\section{eLOCC in quantum critical systems}

In this section we use the above method to study some typical
quantum critical systems. For a Hamiltonian $H(\lambda)$ with a
critical point $\lambda_c$, we change the parameter $\lambda$
gradually from phase I to phase II. For each given $\lambda$, we
calculate and plot the ground state $S_{\alpha}$ with respect to
$\alpha$. We expect to see the two cases described in Table I
emergence.

We first consider the one dimensional spin $1/2$  $XY$ chain with
the Hamiltonian
\begin{eqnarray}
H=-\sum_{i=1}^N[(1+\gamma)\sigma_i^x\sigma_{i+1}^x+(1-\gamma)\sigma_i^y\sigma_{i+1}^y+h\sigma_i^z],
\end{eqnarray}
where $\sigma_i^{x,y,z}$ stand for the Pauli matrices. $\gamma$ is
coupling parameter and $h$ is field parameter. We can focus on the
positive valued $\gamma$ and $h$ space because of the symmetry of
the Hamiltonian. This model and its generalizations have been
studied extensively \cite{xy quench disorder xy fidelity Korepin}.
Fig. 2 shows the phase diagram of this model. This two dimensional
phase diagram is a little bit complicated. In order to make a
clearer description of the eLOCC proposal, we first employ the
transverse field $Ising$ model, which is a special case of the $XY$
model to show how this method works. We can obtain the transverse
field Ising model from the XY
  Hamiltonian by setting $\gamma=1$, $h=2g$ and removing the
  unimportant global factor $2$ for each components as
\begin{eqnarray}
H_I=-\sum_{i=1}^N(\sigma_i^x\sigma_{i+1}^x+g\sigma_i^z).
\end{eqnarray}
A second order quantum phase transition takes place at $g=1$. When
$g\neq1$, there is a energy gap between the ground state and the
first excited state, and this gap vanishes when $g=1$ \cite{quantum
phase transition}.

\begin{figure}
\includegraphics[height=4cm,width=5cm]{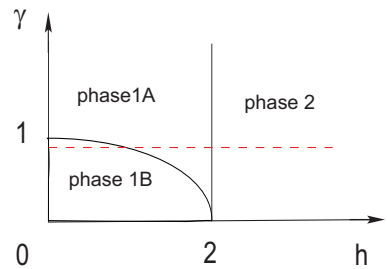}
\caption{\label{fig2:epsart} (Color online). Phase diagram of the
$XY$ model. The phase transition from phase $1A$ or phase $1B$ to
phase $2$ referred to as the $Ising$ transition is driven by the
transverse magnetic field $h$, and the phase boundary $h=2$ is a
critical line. Phase $1$ has two distinct regions $A$ and $B$. The
boundary $\gamma^2+(h/2)^2=1$ is not a critical line. But the
properties of this two regions are different. We consider the red
dash line $\gamma=\sqrt 3/2$ in the phase diagram. Thus, it
intercepts with the two boundaries at $h=1$ and $h=2$, respectively.
}
\end{figure}

We calculate the R\'{e}nyi entropies of the ground states with the
parameter $g$ varying from $0.5$ to $1.5$ and plot them in Fig. 3.
Here the system size $N=10$ and periodical boundary condition is
assumed, i.e., $\sigma_{N+1}^{x,y}=\sigma_1^{x,y}$. To calculate the
R\'{e}nyi entropy we need to know the ground state, which is
obtained by diagonalizing the whole Hamiltonian using Matlab.
Although the system size which can be calculated is relatively
small, the advantage of directly diagonalizing the whole Hamiltonian
is the high accuracy. Other numerical methods, such as Lanczos
algorithm, DMRG and so on are worth generalizing in this proposal,
but as we are conceiving a new proposal to investigate the quantum
critical point rather than developing a different numeric technique
within the framework of those already known proposals, we focus on
the basic diagonalization  method, and the application of other
numeric methods are not within the scope of the present paper.

The results have shown that these states can be classified into
three distinct groups: In group I (blue line) $g<1$ ; In group II
(red line) $g$ is at transition point and in group III (black line)
$g>1$ . Group I and III are two phases and group II is the phase
transition region, which is the boundary of I and III. Fig. 3
clearly shows that in group I R\'{e}nyi entropies for different
ground states intercept with each other. While in group III the
R\'{e}nyi entropy of different states never intercepts with each
other. Apart from that the R\'{e}nyi entropy from different phases
will intercept.

\begin{figure}
\includegraphics[height=6cm,width=8cm]{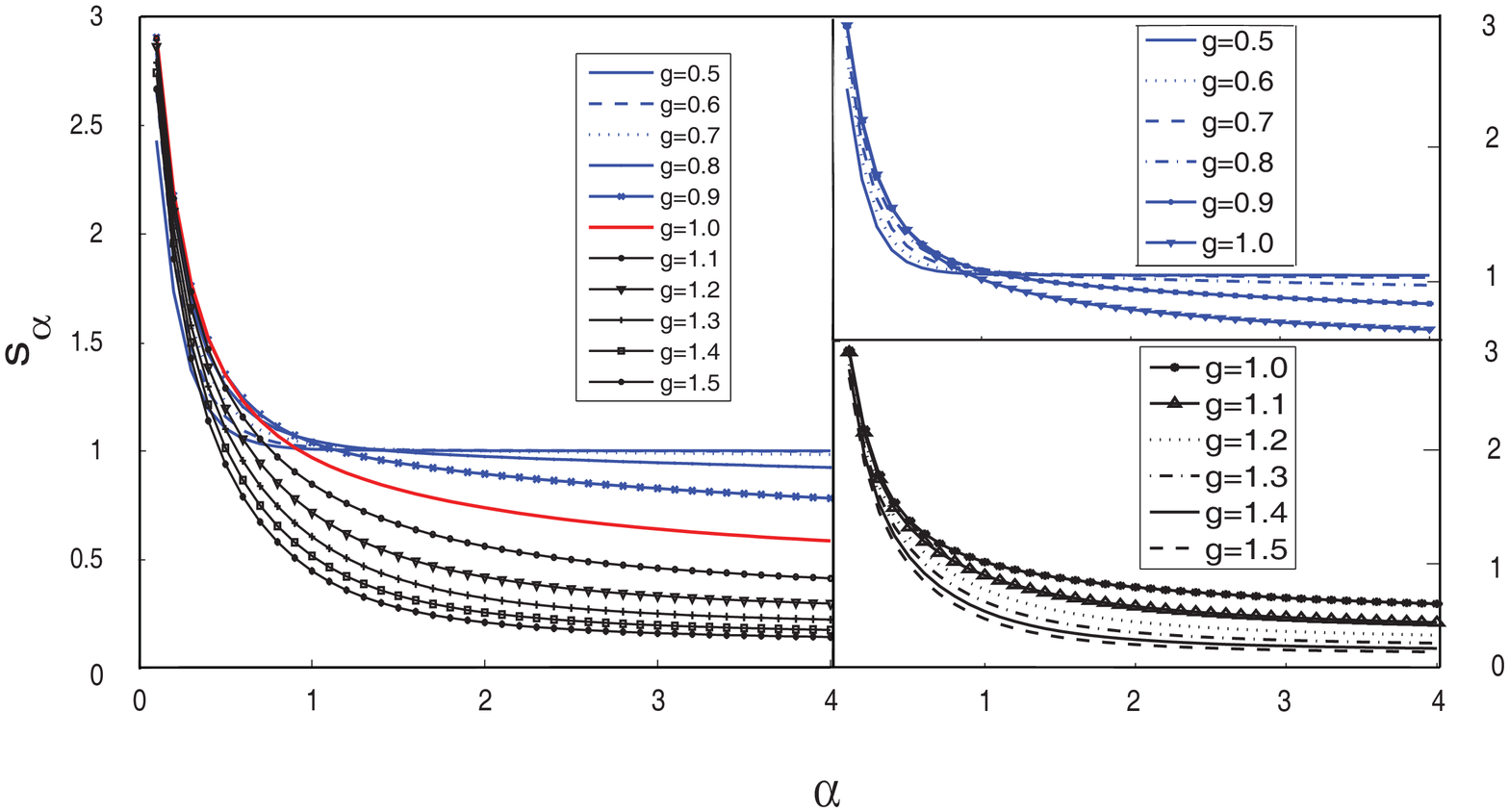}
\caption{\label{fig:epsart} (Color online). R\'{e}nyi entropy for
the ground state of the transverse field Ising  model versus the
parameter $\alpha$. The total site number $N$ is taken to be $10$,
and the spins are cut into two blocks with $5$ sites respectively.
Periodical boundary condition is assumed. The blue lines are for the
ground states with $g=0.5,0.6,0.7,0.8,0.9$, and the black lines are
for the ground states with $g=1.1,1.2,1.3,1.4,1.5$,respectively. The
red line is $g=1$. }
\end{figure}

These results mean that although in region I and III the model are
both gapped their ground states convertibility by eLOCC are quite
different: In the ferromagnetic phase where $g<1$, the ground states
cannot convert to each other because their R\'{e}nyi entropy
intercepts; while in the paramagnetic phase where $g>1$, the ground
states can convert by eLOCC because their R\'{e}nyi entropies never
intercept. In addition, states from different regions cannot convert
to each other by eLOCC.  We can conclude from here that the phase
transition has its global signature, and the long range correlations
which influence the eLOCC convertibility must play a fundamental
role.

Due to the resolving limit of human eyes, we can illustrate the
results better by directly comparing the R\'{e}nyi entropy data in a
table instead of the $S_{\alpha}$ versus $\alpha$ figure, please see
table II. It shows whether any two ground states intercept or not.
For example in the second row of Table II, we find that the ground
state of $g=0.94$ intercepts with those of $g=0.95, 0.96$,..., at
$\alpha =0.6, 0.5$,.... For the sack of space limit and clarity, we
only present the segment of g taking from $0.94$ to$1.04$. Notice
that the diagonal elements are always 'N', which means there is no
interception between the corresponding two states, because they
stand for two same states must completely overlap but not intercept.
By Table II, we find the phase transition lies in $0.98\leq
g\leq1.00$. We can go on investigating this phase transition more
accurately by the same method and we list the result here: when the
accuracy (g step) is $0.001$ the critical region obtained by this
method is $0.987\leq g\leq0.989$, and  the critical region is
$0.9883\leq g\leq0.9885$ with the accuracy of $0.0001$. We can see
from the results that the interval becomes smaller as the varying
step of the parameter $g$ becomes more accurate. Moreover, the
critical value we obtain is not exactly $1$ because we only
calculate a finite chain with $10$ spins. To get rid of the finite
size effect, we give the scaling analysis in Fig.4. When
$N\rightarrow\infty$, the phase transition point obtained by this
proposal tends to $0.9940$. This Ising model corresponds to the
first case of the proposal described in the previous section, and
Table II corresponds to the left type of Table I.

\begin{table}
\caption{\label{tab2} Interception table of transverse field Ising
model near the critical point. For clearance, we cut the table into
separate parts, $g\le 0.98$, $g\ge 1$ and $0.98\le g\le 1$.}
\begin{scriptsize}  
\begin{tabular}{|c|c|c|c|c|c||c||c|c|c|c|c|}
\hline
  g & 0.94&0.95&0.96&0.97&0.98&0.99&1.00&1.01&1.02&1.03&1.04\\
  \hline

 0.94& N&  0.6 &0.5 &0.5 &0.5 &0.4 &0.4 &0.3 &0.3 &0.2 &N\\
  \hline
 0.95 &0.6 &N  &0.5 &0.5 &0.4 &0.4 &0.3 &0.3 &0.2& N& N\\
  \hline
  0.96 &0.5 &0.5 &N  &0.4 &0.4 &0.3 &0.3 &0.2  &N  &N
  &N\\
\hline
  0.97 &0.5 &0.5 &0.4 &N  &0.3 &0.3 &0.2   &N &N  &N&  N\\
  \hline
  0.98 &0.5 &0.4 &0.4 &0.3 &N  &0.2   &N  &N  &N  &N&  N\\
  \hline
  \hline
0.99 &0.4 &0.4 &0.3 &0.3 &0.2  &N &N &N &N& N &N\\
  \hline
  \hline
  1.00 &0.4 &0.3 &0.3 &0.2 &N    &N  &N  &N  &N  &N&
  N\\
  \hline
  1.01 &0.3 &0.3 &0.2&N  &N  &N    &N &N &N &N  &N\\
  \hline
  1.02 &0.3 &0.2 &N  &N   &N  &N  &N  &N  &N  &N
  &N\\
  \hline
  1.03 &0.2 &N  &N  &N  &N  &N  &N  &N  &N  &N
  &N\\
  \hline
  1.04 &N  &N  &N    &N  &N  &N  &N  &N  &N  &N
  &N\\
  \hline

\end{tabular}
\end{scriptsize}
\end{table}

\begin{figure}
\includegraphics[height=4cm,width=\linewidth]{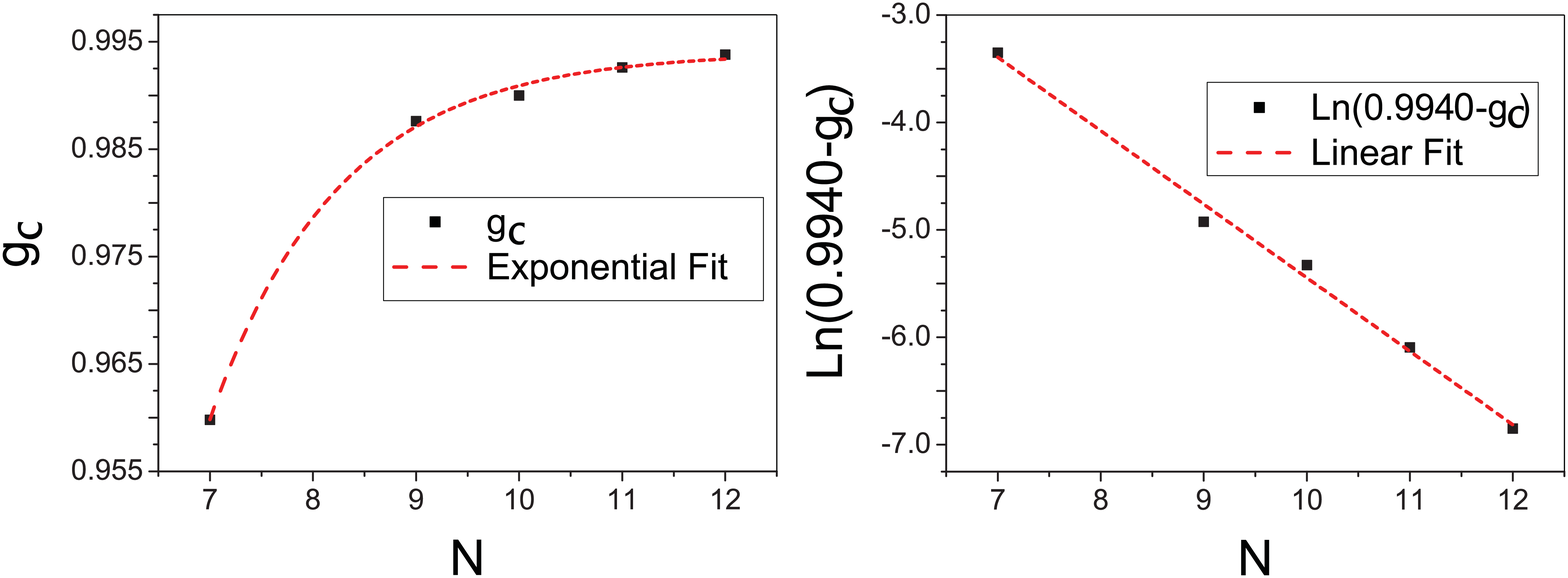}
\caption{\label{fig4:epsart} (Color online).The finite size scaling
analysis of Ising model. In the thermodynamic limit, the critical
point labeled as $g_c$ obtained by our method tends to $0.9940$ with
the accuracy of $0.0001$. The data can be fitted as
$g_c=-9.149e^{-N/1.2522}+0.9940$.}
\end{figure}

So far we have shown the role of R\'{e}nyi entropy in detecting the
critical point by investigating the eLOCC convertibility between
different ground states. Then we find the eLOCC convertibility
between the ground state and the first excited state also gives good
discrimination of different phases. In the ferromagnetic phase where
$g<1$, the R\'{e}nyi entropy  of the ground state and the
corresponding first excited state intercepts, while in the
paramagnetic phase where $g>1$, the two R\'{e}nyi entropies have no
interception. We show this in Fig. 5. Moreover the difference of the
R\'{e}nyi entropy in the large $\alpha$ limit between the excited
state and the ground state becomes larger with the increasing of the
energy gap in the paramagnetic phase.

\begin{figure}[h]
\includegraphics[height=3cm,width=\linewidth]{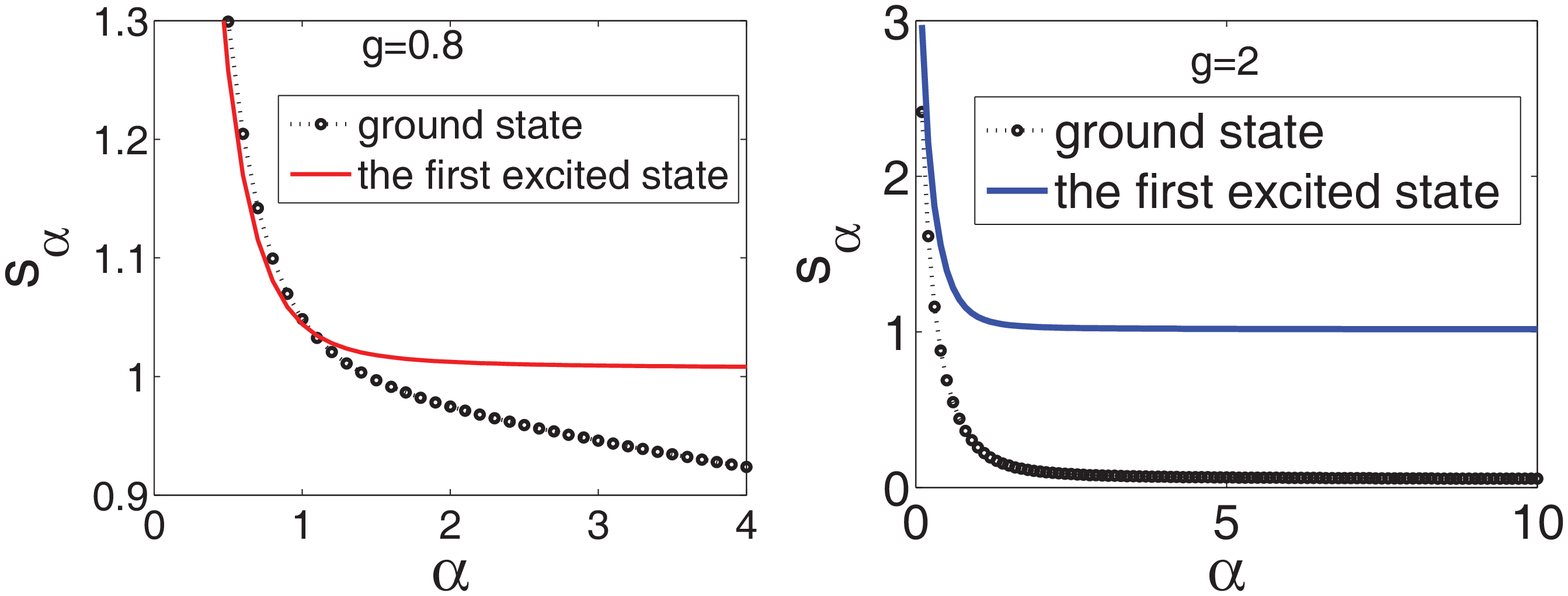}
\caption{\label{fig6:epsart} (Color online). R\'{e}nyi entropies of
ground state and the first excited state. The dash lines are for the
ground states, and the solid lines are the first excited states.}
\end{figure}

\begin{table}[h]
\caption{\label{tab3}Interception table of the XY model near the
first boundary $h=1$. }
\begin{scriptsize}  
\begin{tabular}{|c|c|c|c|c|c|c|c|c|c|c|c|c|c|c|c|}
\hline h&  0.5& 0.6& 0.7 &0.8 &0.9 &1   &1.1 &1.2 &1.3 &1.4 &1.5      \\
\hline 0.5   &N&N&N&N&N&N&N&N&  0.3& 1.5 &1.4\\
\hline 0.6   &N&N&N&N&N&N&N&N&  0.2& 1.5 &1.5\\
\hline 0.7   &N&N&N&N&N&N&N&  0.5& 1.6 &1.6 &1.5 \\
\hline 0.8   &N&N&N&N&N&N&N&  1.6 &1.8 &1.6 &1.5\\
\hline 0.9   &N&N&N&N&N&N&  1.4& 2   &1.9 &1.7 &1.5\\
\hline 1     &N&N&N&N&N&N&  2.2 &2.2 &1.9 &1.7 &1.5\\
\hline 1.1   &N&N&N&N  &1.4 &2.2 &N  &2.2 &1.9 &1.7 &1.5\\
\hline 1.2   &N&N&  0.5 &1.6 &2   &2.2 &2.2 &N  &1.8 &1.7 &1.5\\
\hline 1.3 & 0.3 &0.2 &1.6 &1.8 &1.9 &1.9 &1.9 &1.8 &N  &1.6 &1.5\\
\hline 1.4 & 1.5 &1.5 &1.6 &1.6 &1.7 &1.7 &1.7 &1.7 &1.6 &N  &1.4\\
\hline 1.5 & 1.4 &1.5 &1.5 &1.5 &1.5 &1.5 &1.5 &1.5 &1.5 &1.4&N\\
  \hline
\end{tabular}
\end{scriptsize}
\end{table}

\begin{table}[h]
\caption{\label{tab4}Interception table of the XY model near the
 critical point $h=2$. }
\begin{scriptsize}  
\begin{tabular}{|c|c|c|c|c|c|c|c|c|c|c|c|c|c|c|c|}
\hline h&  1.6 &1.7 &1.8 &1.9 &2   &2.1 &2.2 &2.3 &2.4 &2.5 &2.6      \\
\hline 1.6 &N  &1.1 &1   &0.9 &0.8 &0.7 &0.6 &0.5 &0.4 &0.3 &N\\
\hline 1.7 &1.1 &N  &0.9 &0.8 &0.7 &0.6 &0.4 &0.2 &N&N&N\\
\hline 1.8 &1   &0.9 &N  &0.7 &0.6 &0.4 &0.3 &N&N&N&N \\
\hline 1.9 &0.9 &0.8 &0.7 &N  &0.4 &0.2 &N&N&N&N&N\\
\hline 2   &0.8 &0.7 &0.6 &0.4 &N&N&N&N&N&N&N\\
\hline 2.1 &0.7 &0.6 &0.4 &0.2 &N&N&N&N&N&N&N\\
\hline 2.2 &0.6 &0.4 &0.3 &N&N&N&N&N&N&N&N\\
\hline 2.3 &0.5 &0.2 &N&N&N&N&N&N&N&N&N\\
\hline 2.4 &0.4 &N&N&N&N&N&N&N&N&N&N\\
\hline 2.5 &0.3 &N&N&N&N&N&N&N&N&N&N\\
\hline 2.6 &N&N&N&N&N&N&N&N&N&N&N  \\
  \hline
\end{tabular}
\end{scriptsize}
\end{table}

In order to study the general case of $XY$ model we set
$\gamma=\sqrt3/2$ and varying $h$ from $0$ to $3$, see the red dash
line in Fig.2.
 We can use the eLOCC proposal to determine the critical point at $h=2$ as well as the boundary at $h=1$.
 Here we just list the results: considering the eLOCC convertibility
between ground states, there is no interception in region $1B$, but
every two ground states in  region $1A$ intercept, and then in
phase$2$ there is no interception again, please see table III and
IV. The boundary at $h=1$ and the critical point at $h=2$ also
correspond to the first case introduced in the previous section.
Table III and IV are corresponding to the left type in Table I.
There are interceptions between region $1B$ and phase$2$. For the
case of total lattice number $N=10$, we detect the first boundary
lies in the range $0.999\leq h\leq1.000$ and the critical point is
$2.010\leq h\leq2.012$ with the accuracy of $0.001$.

Next, we study  one dimensional XXZ model with the Hamiltonian
\begin{eqnarray}
H_{XXZ}=\sum_i\sigma_i^x\sigma_{i+1}^x+\sigma_i^y\sigma_{i+1}^y+\Delta
\sigma_i^z\sigma_{i+1}^z,
\end{eqnarray}
where $\Delta$ is the anisotropic parameter. There are two phase
transition points: $\Delta=-1$ corresponds to a first order phase
transition, and $\Delta=1$ is a continuous phase transition of
infinite order \cite{CNYang}. In particular, the phase transition at
$\Delta=1$ is a Kosterlitz-Thouless like transition, which can be
described by the correlation length divergency but without long
range order \cite{yang}. We focus on identifying the critical point
$\Delta=1$ by the eLOCC proposal. Here we use the same method as we
did in the Ising model and XY model to get the ground state
R\'{e}nyi entropy, i.e., diagonalizing the whole Hamiltonian to
obtain all the eigenstates, then we select the ground state to
calculate the eigenvalues of reduced density matrix to obtain the
R\'{e}nyi entropy.

Table V shows the interceptions near $\Delta =1$. We can see that
each state in either region $\Delta \ge 1.0$ or $\Delta \le 1.0$
never intercepts with any of the states in the same region, but
intercepts with at least one state from the other region. Therefore
this model corresponds to the second case of the proposal introduced
in the previous section. Thus, the critical region can be determined
as $0.9\le \Delta \le 1.1$. By narrowing the varying step of
$\Delta$, this critical point can be detected more accurately.
Therefore, the eLOCC proposal also works well for this infinite
order phase transition in $XXZ$ spin chain.

We can find that result of $XXZ$ model resembles the right pattern
of Table I, i.e., the ground states do not intercept with the states
form the same phase, but intercepts the states from the other phase.

\begin{table}[h]
\caption{\label{tab1} Interception table of $XXZ$ model. For
clearance, we cut the table into separate parts, $\Delta\le 0.9$,
$\Delta\ge 1.1$ and $0.9\le \Delta\le 1.1$. It is a $10$ sites chain
with comb like partition, i.e., the odd numbered sites belong to
part $A$. }
\begin{scriptsize}  
\begin{tabular}{|c|c|c|c|c|c|c||c||c|c|c|c|c|c|}
\hline
  $\Delta$ &0.4& 0.5& 0.6 &0.7 &0.8 &0.9 &1.0   &1.1 &1.2 &1.3 &1.4
   &1.5&1.6\\
   \hline
    0.4 &N &N &N &N &N &N &N &N  &N &N &N &5.4 &1.1\\
\hline
 0.5 &N&N &N &N &N &N &N &N &N  &N &N  &1&N\\
 \hline
 0.6 &N& N &N &N &N &N &N &N &N  &N &1  &N&N\\
\hline
 0.7&N &N &N &N &N &N &N &N &N    &0.9 &N &N&N\\
 \hline
 0.8&N &N &N &N &N &N &N &N    &0.9 &N &N &N&N\\
\hline
 0.9 &N&N &N &N &N &N &N  &0.9 &N &N &N &N&N\\
\hline
 \hline
 1.0 &N&N &N &N &N &N &N &N &N  &N &N &N&N\\
\hline \hline
 1.1 &N&N &N &N &N  &0.9 &N &N &N &N &N &N&N\\
\hline
 1.2 &N&N &N &N  &0.9 &N &N &N &N &N &N &N&N\\
\hline
 1.3&N &N &N  &0.9 &N &N &N &N &N &N &N &N&N\\
\hline
 1.4 &N&N  &1   &N &N &N &N &N &N &N &N  &N&N\\
\hline
 1.5 &5.4&1   &N &N &N &N &N &N &N &N  &N &N&N\\
 \hline
  1.6 &1.1&N &N &N &N &N &N &N &N  &N &N &N&N\\
\hline
\end{tabular}
\end{scriptsize}
\end{table}

\section{Conclusions}

 In this paper, we analyzed the R\'{e}nyi entropy and the eLOCC convertibility
 in quantum critical systems. We developed a new proposal to describe the eLOCC conversion
properties in quantum critical systems by examining the R\'{e}nyi
entropy interception, which shows the disability of eLOCC
conversion. We applied this proposal to the study of several typical
quantum critical systems, e.g. one dimensional transverse field
Ising model, $XY$ model and XXZ model.
 The results
showed that: the eLOCC convertibility changes at critical points.
For the Ising phase transition, eLOCC convertibility in the
paramagnetic phase is stronger than that in the ferromagnetic phase
in two ways: (i) any two ground states in paramagnetic phase can
convert by eLOCC but those in ferromagnetic phase cannot; (ii) each
first excited state in paramagnetic phase can be converted to their
ground states by eLOCC while states in ferromagnetic phase cannot.
For the XY model with two dimensional phase diagram, the critical
points can be determined via this method at very high accuracy. The
boundary between region $1A$ and $1B$ can be detected as well. For
the XXZ model the infinite order phase transition at $\Delta=1$ can
also be determined by this method while the pattern of the
interception table is different from the  second order quantum phase
transitions in Ising model and XY model.

In fact, as is shown in Table I, the R\'{e}nyi entropy inception
table can be divided into four blocks. The two anti-diagonal blocks
represent the comparison between ground states from two different
phases, and it is not surprising that the two blocks are crossings
which means that the ground states from the two phases are too
different to convert by local operations. The two diagonal blocks
represent the comparison between the ground states from the same
phase. We find two possible cases in Table I, which are two
extremes. Case (i) has the most crossings in the table, while case
(ii) has the least crossings. As a matter of fact, The crossings in
the table reflect the degree of how hard it is to convert the
different ground states, which can be served as a measure of the
difference between the two phases. It is quite interesting to find
that this observation is amazingly consistent with the orders of the
phase transitions in the two example models. In case (i), i.e., the
second order quantum phase transition which is the lowest order that
quantum critical phenomena can exist, we find the pattern with the
most crossings, while in case (ii), i.e., the infinite order quantum
phase transition, we find the pattern with the least crossings.

The eLOCC proposal may help further understanding the complicated
phenomena of quantum critical systems. This paper would initiate
extensive studies of quantum phase transitions from the brand new
perspective of local convertibility. This simple but effective
method is worth (a) generalizing to study finite temperature phase
transitions (b) generalizing based on the majorization scheme
 \cite{Nielsen} and (c) applying to other systems.

{\label{sec:level1}} \emph{Acknowledgement}. ---One of the authors
J. Cui thanks the helpful discussion with Zhi-Hao Xu and Zhao Liu.
This work is supported by NSFC grant and the National Program for
Basic Research of MOST (``973'' program).

\newpage

\end{document}